\documentstyle[12pt,qqa4lart]{article}
\input tcilatex

\QQQ{Language}{
American English
}

\begin{document}

\author{Lu-Ming Duan$^{*}$ and Guang-Can Guo\thanks{%
Electronic address: gcguo@sunlx06.nsc.ustc.edu.en} \\
Department of Physics and Nonlinear Science Center,\\
University of Science and Technology of China,\\
Hefei, Anhui 230026, People's  Republic of China}
\title{Perturbative expansions for the fidelities \\
and spatially-correlated dissipation of quantum bits }
\date{}
\maketitle

\begin{abstract}
\baselineskip 24ptWe construct generally applicable short-time perturbative
expansions for some fidelities, such as the input-output fidelity, the
entanglement fidelity, and the average fidelity. Successive terms of these
expansions yield characteristic times for the damping of the fidelities
involving successive powers of the Hamiltonian. The second-order results,
which represent the damping rates of the fidelities, are extensively
discussed. As an interesting application of these expansions, we use them to
study the spatially-correlated dissipation of quantum bits. Spatial
correlations in the dissipation are described by a correlation function.
Explicit conditions are derived for independent decoherence and for
collective decoherence.\\

\end{abstract}

\newpage\ \baselineskip 24ptIn various fields of physics, the study of open
quantum systems plays an important role. For example, decoherence was
recognized as a major problem in realizing quantum computation [1,2].
Decoherence in quantum computers mainly results from the coupling of the
quantum bit (qubits) to the environment. In general, It is not practical to
look for exact solutions of these complicated systems, which consists of
many spatially-correlated qubits interacting with a reservoir.

In this paper, we propose a perturbative approach to the study of open
quantum systems. We construct generally applicable short-time perturbative
expansions for some fidelities, such as the input-output fidelity, the
entanglement fidelity, and the average fidelity. A similar perturbative
expansion for coherence loss has been proposed in a recent paper [3], where
the coherence loss is measured by the quantity $\delta \left( t\right)
=tr\left[ \rho \left( t\right) -\rho ^2\left( t\right) \right] $, in which $%
\rho \left( t\right) $ indicates the reduced density of the system. Since $%
\delta \left( t\right) $ has no direct physical meanings, in this paper, we
choose the fidelities rather than $\delta \left( t\right) $ as measures of
decoherence. The fidelities are important quantities and have been widely
used in quantum coding theory [4-7]. Through the perturbative expansions, we
can get some general relations between the fidelities, such as the
demonstration that the entanglement fidelity decays more rapidly than the
average fidelity.

As an interesting application of this perturbative approach, we use it to
study spatially-correlated dissipation of the qubits. The latter question is
of great practical importance in designing quantum computers. Spatial
correlations in pure dephasing of the qubits have been analyzed in Refs. [8]
and [9]. However, the calculations of spatial correlations in general
dissipation of the qubits, including the dephasing and the relaxation, are
far more involved. This question is hard to solve either by any existing
exact approaches or by solving the master equation [10]. Fortunately, here
we show, through the perturbative expansions, spatial correlations in
general dissipation of the qubits can be easily analyzed. Spatial
correlations are described by a correlation function. From this result, we
derive explicit conditions for independent decoherence and for collective
decoherence, which are two important ideal circumstances in dissipation of
the qubits.

\section{Perturbative expansions for the fidelities}

We consider an open quantum system, the total Hamiltonian of which is
expressed as 
\begin{equation}
\label{1}H_T=H_0+H_I+H_{env},
\end{equation}
where $H_0$ and $H_{env}$ indicate the free Hamiltonians of the system and
of the environment, respectively. $H_I$ is the interaction Hamiltonian
between the system and the environment. First, we suppose that the system is
initially in a pure state $\left| \Psi _0\right\rangle $. If there is no
coupling between the system and the environment, at time $t$ the system is
in the state 
\begin{equation}
\label{2}\left| \Psi \left( t\right) \right\rangle =e^{-iH_0t/\hbar }\left|
\Psi _0\right\rangle =U_0\left( t\right) \left| \Psi _0\right\rangle .
\end{equation}
But in reality, coupling of the system to the environment is inevitable. So
at time $t$ the system is in fact described by the reduced density 
\begin{equation}
\label{3}\rho \left( t\right) =tr_{env}\left[ \exp \left( -iH_Tt/\hbar
\right) \rho _{env}\left( 0\right) \otimes \left| \Psi _0\right\rangle
\left\langle \Psi _0\right| \exp \left( iH_Tt/\hbar \right) \right] ,
\end{equation}
where $\rho _{env}\left( 0\right) $ is the initial density of the
environment. Decoherence of the system due to this inevitable coupling can
therefore be measured by the fidelity between the state (2) and (3), which
has the form 
\begin{equation}
\label{4}F\left( t\right) =\left\langle \Psi _0\right| U_0^{+}\left(
t\right) \rho \left( t\right) U_0\left( t\right) \left| \Psi _0\right\rangle
.
\end{equation}
The input-output fidelity is first defined in Ref. [4]. Here the definition
is slightly modified to Eq. (4) to make it more suitable for measuring
decoherence of the system.

In general, the reduced density $\rho \left( t\right) $ in Eq. (3) is hard
to calculate. So there is difficulty to obtain the fidelity (4).
Fortunately, in practice, the short-time behaviors of the system under
dissipation are of most interests. Hence we do not need to exactly calculate
the fidelity (4), but expand it into a short-time power series 
\begin{equation}
\label{5}F\left( t\right) =1-\frac t{\tau _1}-\frac{t^2}{\tau _2^2}-\cdots . 
\end{equation}
Following Eqs. (2), (3) and (4), it is not difficult to obtain the expansion
coefficients. Up to order $t^2$ they read explicitly 
\begin{equation}
\label{6}\frac 1{\tau _1}=0 
\end{equation}
\begin{equation}
\label{7}
\begin{array}{c}
\frac{\hbar ^2}{2\tau _2^2}=\left\langle \Psi _0\right| tr_{env}\left\{
\left[ H_I,\left[ H_I,\rho _{env}\left( 0\right) \otimes \left| \Psi
_0\right\rangle \left\langle \Psi _0\right| \right] \right] \right. \\  \\ 
+\left. \left[ \left[ H_I,H_0\right] ,\rho _{env}\left( 0\right) \otimes
\left| \Psi _0\right\rangle \left\langle \Psi _0\right| \right] \right\}
\left| \Psi _0\right\rangle \\  
\\ 
=\left\langle H_I^2\right\rangle _{s,env}-\left\langle \left\langle
H_I\right\rangle _s^2\right\rangle _{env}, 
\end{array}
\end{equation}
where the symbol $\left\langle \cdots \right\rangle _{s,env}$ stands for the
average value over the system and the environment, i.e., $\left\langle
H_I\right\rangle _s=\left\langle \Psi _0\right| H_I$ $\left| \Psi
_0\right\rangle ,$ and $\left\langle H_I\right\rangle
_{s,env}=tr_{env}\left[ \rho _{env}\left( 0\right) \left\langle
H_I\right\rangle _s\right] $. It is obvious that the expansion coefficients
in Eq. (5) furnish characteristic times $\tau _n$ associated with
decoherence processes involving $H_I$ to order $n$. The second-order
coefficient $\frac 1{\tau _2^2}$ is of special interest. It is the
coefficient of the first non-trivial term in the expansion. Under the
short-time approximation, $\frac 1{\tau _2}$ measures the damping rate of
the fidelity. In most cases, it is sufficient to consider the expansion up
to order $t^2$.

In the above, we have assumed that the system is initially in a pure state.
The input-output fidelity (4) does not apply for the mixed states. For a
mixed input state of the system, there are some fidelities defined, such as
the entanglement fidelity and the average fidelity [5]. These two fidelities
are widely used in quantum coding theory. Suppose $\left| \Psi
_{rs}\right\rangle $ is a purification of the density $\rho _s$, i.e., the
state $\left| \Psi _{rs}\right\rangle $ satisfies $tr_r\left( \left| \Psi
_{rs}\right\rangle \left\langle \Psi _{rs}\right| \right) =\rho _s$, where
the symbol $r$ denotes an ancillary system. Similar to Eq. (4), the
entanglement fidelity $F_e$ is defined as 
\begin{equation}
\label{8}
\begin{array}{c}
F_e\left( t\right) =\left\langle \Psi _{rs}\right| U_0^{+}\left( t\right)
tr_{env}\left[ \exp \left( -iH_Tt/\hbar \right) \rho _{env}\left( 0\right)
\right.  \\  
\\ 
\left. \otimes \left| \Psi _{rs}\right\rangle \left\langle \Psi _{rs}\right|
\exp \left( iH_Tt/\hbar \right) \right] U_0\left( t\right) \left| \Psi
_{rs}\right\rangle  \\  
\\ 
=1-\frac t{\tau _{1e}}-\frac{t^2}{\tau _{2e}^2}-\cdots ,
\end{array}
\end{equation}
where $F_e\left( t\right) $ is furthermore subjected to a short-time power
series expansion. In Ref. [5], it has been proven that the entanglement
fidelity defined above is an intrinsic quantity of the system $s$, i.e., it
does not depend on the specific purification $\left| \Psi _{rs}\right\rangle 
$. From Eq. (5), we easily obtain 
\begin{equation}
\label{9}\frac 1{\tau _{1e}}=0
\end{equation}
\begin{equation}
\label{10}\frac{\hbar ^2}{2\tau _{2e}^2}=\left\langle H_I^2\right\rangle
_{s,env}-\left\langle \left\langle H_I\right\rangle _s^2\right\rangle _{env}.
\end{equation}
Comparing Eqs. (6) (7) with Eqs. (9) (10), we see, the characteristic times
for the entanglement fidelity have the same forms as those for the
input-output fidelity. Only change is that the symbol $\left\langle
H_I\right\rangle _s$ now means $\left\langle H_I\right\rangle _s=tr_s\left(
\rho _sH_I\right) $. So the entanglement fidelity is a natural extension of
the input-output fidelity to include the mixed input states. Eqs. (9) and
(10) also show, the characteristic times for the entanglement fidelity are
all intrinsic properties of the system $s$.

The initial density of the system can be expressed as a mixture of pure
states, i.e., 
\begin{equation}
\label{11}\rho _s=\stackunder{i}{\sum }p_i\left| \Psi _i\right\rangle
\left\langle \Psi _i\right| ,
\end{equation}
where $p_i$ satisfy $\stackunder{i}{\sum }p_i=1$. The average fidelity $F_a$
is defined as 
\begin{equation}
\label{12}
\begin{array}{c}
F_a\left( t\right) =
\stackunder{i}{\sum }p_iF\left( \left| \Psi _i\right\rangle \right)  \\  \\ 
=1-\frac t{\tau _{1a}}-\frac{t^2}{\tau _{2a}^2}-\cdots ,
\end{array}
\end{equation}
where $F\left( \left| \Psi _i\right\rangle \right) $ denotes the
input-output fidelity for the pure state $\left| \Psi _i\right\rangle $.
Since the expression (11) for the density $\rho _s$ is not unique, unlike
the entanglement fidelity, the average fidelity is not solely defined for a
definite $\rho _s$. From Eqs. (12) and (4), we get the characteristic times
for the average fidelity 
\begin{equation}
\label{13}\frac 1{\tau _{1a}}=0,
\end{equation}
\begin{equation}
\label{14}\frac{\hbar ^2}{2\tau _{2a}^2}=\left\langle H_I^2\right\rangle
_{s,env}-\left\langle \stackunder{i}{\sum }p_i\left\langle H_I\right\rangle
_i^2\right\rangle _{env},
\end{equation}
where the symbol $\left\langle H_I\right\rangle _i$ indicates $\left\langle
\Psi _i\right| H_I$ $\left| \Psi _i\right\rangle $. Comparing Eq. (10) with
Eq. (14), we get an interesting inequality. For any interaction Hamiltonians 
$H_I$, there is the operator inequality 
\begin{equation}
\label{15}\stackunder{i}{\sum }p_i\left\langle H_I\right\rangle _i^2\geq
\left( \stackunder{i}{\sum }p_i\left\langle H_I\right\rangle _i\right)
^2=\left\langle H_I\right\rangle _s^2.
\end{equation}
Hence Eqs. (10 ) and (14) yield 
\begin{equation}
\label{16}\frac 1{\tau _{2e}}\geq \frac 1{\tau _{2a}},
\end{equation}
which suggests, the entanglement fidelity decays more rapidly than any
average fidelities. In Ref. [5] it has been proven that the entanglement
fidelity is always less than the average fidelity. Here we show this fact
from another aspect.

\section{Spatially-correlated dissipation of the qubits}

As an interesting application of the perturbative approach to open quantum
systems, we consider a practical question: decoherence in quantum computers.
This decoherence is due to the inevitable coupling of the qubits to the
external environment. A few papers [1,2,8,9,11-14] have been published on
this subject with some simplifications, such as omitting spatial
correlations in the decoherence [11-14], or omitting relaxation of the
qubits [8,9], or omitting both of them [1,2]. However, in real
circumstances, such as in the ion-trapped quantum computers [15], relaxation
of the qubits has notable contributions to decoherence [11-13]. On the other
hand, spatial correlation properties of decoherence play an important role
in the choice of the decoherence-reducing strategies. For independent
decoherence and for collective decoherence, the decoherence-reducing
strategies are quite different. So here we consider a more practical model
of decoherence. This decoherence is described by a spatially-correlated
amplitude damping, which includes both the dephasing and the relaxation of
the qubits. The environment is modelled by a bath of oscillators with
infinite degrees of freedom and the mode functions of the bath field are
chosen as plane waves. The Hamiltonian describing this decoherence process
has the form 
\begin{equation}
\label{17}H=H_0+\stackunder{l=1}{\stackrel{L}{\sum }}\stackunder{k}{\sum }%
\left[ \hbar g_k\left( e^{-ikr_l}a_k+e^{ikr_l}a_k^{+}\right) \left( \lambda
_1\sigma _l^x+\lambda _2\sigma _l^y\right) \right] +\stackunder{k}{\sum }%
\left( \hbar \omega _ka_k^{+}a_k\right) ,
\end{equation}
where the Pauli operator $\sigma _l$ represents the $l$ qubit and $a_k$ is
the annihilation operator of the both mode $k$. The symbol $r_l$ denotes the
site of the $l$ qubit, and $\lambda _1,\lambda _2$ and $g_k$ are coupling
constants. $H_0$ in Eq. (17) describes the free evolution and the internal
interaction of the qubits. The qubits, whether in memory or in quantum gate
operations, can all be described by the Hamiltonian (17).

The Hamiltonian (17) is very complicated and it is hard to find its exact
solutions. Fortunately, with the perturbative approach developed in the
previous section, this complex system can be easily treated. To analyze the
decoherence, we use the perturbative expansion for the entanglement
fidelity, which returns to the input-output fidelity if the initial state of
the qubits is pure. The environment is supposed in thermal equilibrium,
i.e., the initial density $\rho _{env}\left( 0\right) $ of the bath has the
following form in the coherent representation 
\begin{equation}
\label{18}\rho _{env}\left( 0\right) =\stackunder{k}{\prod }\int d^2\alpha
_k\frac 1{\pi \left\langle N_{\omega _k}\right\rangle }\exp \left( -\frac{%
\left| \alpha _k\right| ^2}{\pi \left\langle N_{\omega _k}\right\rangle }%
\right) \left| \alpha _k\right\rangle \left\langle \alpha _k\right| ,
\end{equation}
where the mean photon (or phonon) number 
\begin{equation}
\label{19}\left\langle N_{\omega _k}\right\rangle =1/\left[ \exp \left( 
\frac{\hbar \omega _k}{k_BT}\right) -1\right] .
\end{equation}
With this density, and substituting the Hamiltonian (17) into Eq. (10), we
obtain the decoherence rate $\frac 1{\tau _{2e}}$ for this system 
\begin{equation}
\label{20}\frac 1{\tau _{2e}^2}=\stackunder{l_1,l_2=1}{\stackrel{L}{\sum }}%
\Omega ^2\left( r_{l_1}-r_{l_2}\right) \left\langle \Delta A_{l_1}\Delta
A_{l_2}\right\rangle _s,
\end{equation}
where the operator 
\begin{equation}
\label{21}A_{l_i}=\lambda _1\sigma _{l_i}^x+\lambda _2\sigma _{l_i}^y\text{ }%
\left( i=1,2\right) 
\end{equation}
and the spatial correlation function 
\begin{equation}
\label{22}\Omega ^2\left( r_{l_1}-r_{l_2}\right) =2\stackunder{k}{\sum }%
\left\{ \left| g_k\right| ^2\cos \left[ k\left( r_{l_1}-r_{l_2}\right)
\right] \coth \left( \frac{\hbar \omega _k}{2k_BT}\right) \right\} .
\end{equation}
Spatial correlation properties of the decoherence is completely determined
by the correlation function $\Omega ^2\left( r_{l_1}-r_{l_2}\right) $.

In the following, we discuss two important ideal circumstances for the
correlation function. Eq. (22) can be rewritten as 
\begin{equation}
\label{23}\Omega ^2\left( r_{l_1}-r_{l_2}\right) =x\stackunder{k}{\sum }%
f\left( k\right) \cos \left[ k\left( r_{l_1}-r_{l_2}\right) \right] ,
\end{equation}
where $f\left( k\right) $ is a normalized distribution satisfying $%
\stackunder{k}{\sum }f\left( k\right) =1$ and $x$ is the normalization
constant 
\begin{equation}
\label{24}x=2\stackunder{k}{\sum }\left| g_k\right| ^2\coth \left( \frac{%
\hbar \omega _k}{2k_BT}\right) .
\end{equation}
The expression of $f\left( k\right) $ is given by comparing (23) with (22).
Its explicit form depends on the coupling coefficient $\left| g_k\right| ^2$%
, whereas the latter is determined by the specific physical model for
quantum computers. As a simplification, here we assume that the distribution 
$f\left( k\right) $ can be approximated by a Gaussian function with an
expectation value $\overline{k}$ and a variance $\Delta k$, respectively.
With this simplification, Eq. (23) reduces to 
\begin{equation}
\label{25}\Omega ^2\left( r_{l_1}-r_{l_2}\right) \approx x\cos \left[ 
\overline{k}\left( r_{l_1}-r_{l_2}\right) \right] \exp \left[ -\frac
12\left( \Delta k\right) ^2\left( r_{l_1}-r_{l_2}\right) ^2\right] .
\end{equation}
Suppose $d$ is distance between the adjacent qubits. If $d$ satisfies 
\begin{equation}
\label{26}\left( \Delta k\right) d>>1,
\end{equation}
Eq. (25) yields 
\begin{equation}
\label{27}\Omega ^2\left( r_{l_1}-r_{l_2}\right) \approx x\delta _{l_1l_2},
\end{equation}
and from Eq. (20) the decoherence rate is simplified to 
\begin{equation}
\label{28}\frac 1{\tau _{2e}^2}=\stackunder{l=1}{\stackrel{L}{\sum }}%
x\left\langle \left( \Delta A_l\right) ^2\right\rangle _s,
\end{equation}
which suggests, the total decoherence rate of $L$ qubits equals the sum of
the decoherence rates of individual qubits. So in this circumstance, the
qubits are decohered independently. Eq. (26) is the condition for
independent decoherence. Most of the existing quantum error correction
schemes are designed to correct for the errors induced by independent
decoherence [16-24].

Apart from the independent decoherence, there is another ideal circumstance
for the spatial correlation function. Suppose there are $2L$ qubits. Two
adjacent qubits make up a qubit-pair. So we have $L$ qubit-pairs. The two
qubits in the $l$ qubit-pair are indicated by $l$ and $l^{^{\prime }}$,
respectively. If distance $d$ between the adjacent qubits satisfies the
condition 
\begin{equation}
\label{29}\overline{k}d<<1\text{ and }\left( \Delta k\right) d<<1,
\end{equation}
form Eq. (25) the spatial correlation function remains a constant for the
two qubits in each qubit-pair. The decoherence rate $\frac 1{\tau _{2e}}$
for $L$ qubit-pairs thus becomes 
\begin{equation}
\label{30}\frac 1{\tau _{2e}^2}=\stackunder{l_1,l_2=1}{\stackrel{L}{\sum }}%
\Omega ^2\left( r_{l_1}-r_{l_2}\right) \left\langle \Delta \left(
A_{l_1}+A_{l_1^{^{\prime }}}\right) \Delta \left( A_{l_2}+A_{l_2^{^{\prime
}}}\right) \right\rangle _s.
\end{equation}
So under the condition (29) the two qubits in each qubit-pair are decohered
collectively. In the collective decoherence, the decoherence rate is
sensitive to the type of the initial states. If the initial state of the
qubit-pairs is a co-eigenstate of all the operators $A_l+A_l^{^{\prime }}$,
form Eq. (30), the second-order decoherence rate $\frac 1{\tau _{2e}}$
reduces to zero. With these states decoherence of the qubits can therefore
be much reduced. The co-eigenstates of the operators $A_l+A_l^{^{\prime }}$
are called the subdecoherent states. In fact, an arbitrary input state of $L$
qubits can be transformed into the corresponding subdecoherent state of $L$
qubit-pairs by the following encoding 
\begin{equation}
\label{31}
\begin{array}{c}
\left| -1\right\rangle \rightarrow \left| -1,+1\right\rangle , \\ 
\left| +1\right\rangle \rightarrow \left| +1,-1\right\rangle ,
\end{array}
\end{equation}
where $\left| -1\right\rangle $ and $\left| +1\right\rangle $ are two
eigenstates of the operator $A_l$. Obviously, the encoded state is a
co-eigenstate of the operators $A_l+A_l^{^{\prime }}$ with the eigenvalue $0$%
. This encoding has been mentioned in [8] and [25] and extended in [26] and
[27] to reduce decoherence in general circumstances. Here we derive the
working condition (29) for this encoding.

In the above, $\overline{k}$ and $\Delta k$ are introduced
phenomenologically. It is assumed that we have no knowledge about the
coupling coefficient $\left| g_k\right| ^2$. There is an interesting case in
which the spatial correlation function can be calculated exactly. Consider
one dimensional quantum computers. In the continuum limit, it can be assumed
that
\begin{equation}
\label{32}\stackunder{k}{\sum }\left| g_k\right| ^2\cdots \propto
\int_0^\infty d\omega _k\omega _ke^{-\omega _k/\omega _c}\cdots 
\end{equation}
where $\omega _c$ is the cut-off frequency whose specific value depends on
the particular nature of the physical qubit under investigation. The form
(32) of $\left| g_k\right| ^2$ was also used in Refs. [1] and [8]. We
consider two circumstances. In the high temperature limit, i.e., $T>>\frac{%
\hbar \omega _c}{k_B}$, the correlation function (22) is simplified to
\begin{equation}
\label{33}
\begin{array}{c}
\Omega ^2\left( r_{l_1}-r_{l_2}\right) \propto 
\frac{4k_BT}\hbar \int_0^\infty d\omega _ke^{-\omega _k/\omega _c}\cos
\left[ \frac{\omega _k}v\left( r_{l_1}-r_{l_2}\right) \right]  \\  \\ 
=\frac{4k_BT}{\hbar \omega _c}\frac{\omega _c^2}{1+\left[ \omega _c\left(
r_{l_1}-r_{l_2}\right) /v\right] ^2},
\end{array}
\end{equation}
where $v$ indicates velocity of the noise field. If distance $d$ between the
adjacent qubits satisfies $d>>v/\omega _c$, Eq. (33) tends to a
delta-function and the qubits are therefore decohered independently. On the
other hand, if $d<<v/\omega _c$, we have $\Omega ^2\left( d\right) =\Omega
^2\left( 0\right) $ and the adjacent qubits are then decohered collectively.
In the low temperature limit, things are much similar. If $T<<\frac{\hbar
\omega _c}{k_B}$, $\coth \left( \frac{\hbar \omega _k}{2k_BT}\right) \approx
1$ and Eq. (22) yields
\begin{equation}
\label{34}
\begin{array}{c}
\Omega ^2\left( r_{l_1}-r_{l_2}\right) \propto 2\int_0^\infty d\omega
_k\omega _ke^{-\omega _k/\omega _c}\cos \left[ 
\frac{\omega _k}v\left( r_{l_1}-r_{l_2}\right) \right]  \\  \\ 
=2\omega _c^2\frac{1-\omega _c^2\left( r_{l_1}-r_{l_2}\right) ^2/v^2}{\left[
1+\omega _c^2\left( r_{l_1}-r_{l_2}\right) ^2/v^2\right] ^2}.
\end{array}
\end{equation}
Eq. (34) suggests, the above conditions for independent decoherence and for
collective decoherence still hold at low temperature. The type of
decoherence is mainly determined by distance between the adjacent qubits and
by the cut-off frequency. Temperature of the environment hardly influences
the decoherence type, though it determines the decoherence rate.

\section{Summary}

In this paper, we develop short-time perturbative expansions for some
widely-used fidelities. From the expansions, we demonstrate some interesting
relations between the fidelities. Perturbative expansions for the fidelities
can be used for studying open quantum systems. As an example, we consider
spatially-correlated dissipation of the qubits in the quantum computer.
Spatial correlations in the dissipation are described by a correlation
function and we successfully derive the explicit conditions for independent
decoherence and for collective decoherence. This example suggests, the
perturbative expansions for the fidelities may be proven as a useful tool
for studying open quantum systems.\\

{\bf Acknowledgment}

This project was supported by the National Natural Science Foundation of
China.

\newpage\

\end{document}